# Work-Function-Dependent Reduction of Transition Metal Nitrides in Hydrogen Environments

Abdul Rehman,* Robbert W. E. van de Kruijs, Wesley T. E. van den Beld, Jacobus M. Sturm, and Marcelo Ackermann

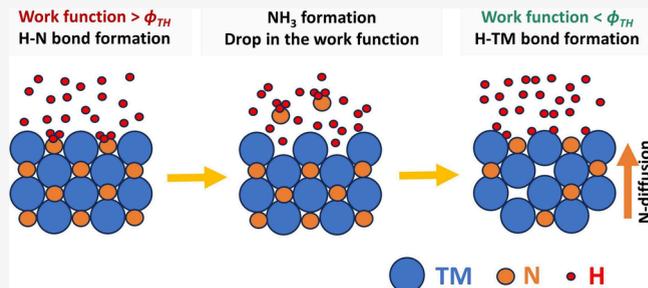

**ABSTRACT:** Amidst the growing importance of hydrogen in a sustainable future, it is crucial to develop coatings that can protect hydrogen-sensitive system components in reactive hydrogen environments. However, the prediction of the chemical stability of materials in hydrogen is not fully understood. In this study, we show that the work function is a key parameter determining the reducibility (i.e., denitridation) of transition metal nitrides (TMNs) in hydrogen radicals (H*) at elevated temperatures. We demonstrate that, when the work function of a TMN system drops below a threshold limit ($\phi_{TH}$), its reduction effectively stops. We propose that this is due to the preferential binding of H* to transition metal (TM) atoms rather than N atoms, which makes the formation of volatile species (NH$_x$) unfavorable. This finding provides a novel perspective for comprehending the interaction of hydrogen with TM compounds and allows prediction of the chemical stability of hydrogen-protective coatings.

Hydrogen is heralded to play a key role in the green energy transition as either a fuel (fusion)[1,2] or energy carrier (green hydrogen).[3,4] Understanding and predicting how hydrogen interacts with materials such as transition metal nitrides (TMNs), which are candidates for hydrogen-protective coatings,[2,5] is, therefore, key to fully utilizing hydrogen's potential for a sustainable future.

The work function is a fundamental parameter of a material system representing the amount of energy required to transfer an electron from the Fermi level to the vacuum level. Literature indicates a strong correlation between a material's work function and how hydrogen interacts with it. For instance, on the basis of simulations, Van de Walle et al. showed that the electronic nature of adsorbed hydrogen, forming a proton (H$^+$) or hydride ion (H$^-$), is directly dependent upon the work function of the surface material for semiconductors and insulators.[6,7] Furthermore, Kura et al. and Saito et al. experimentally demonstrated that hydrogen in TiN$_x$, HfN$_x$, and Zr$_3$N$_{4-\delta}$ adsorbs as H$^-$, forming bonds with transition metal (TM) atoms, proposing that this behavior is due to the materials' low work functions.[8−10]

Van de Walle et al.[6,7] proposed a universal (material-independent) work function threshold ($\phi_{TH}$), at which the formation energies of H$^+$ and H$^-$ are equal. This $\phi_{TH}$ is expected to be independent of both the hydrogen's state (energy) and its concentration, as these factors only influence the formation energy of the reactants, while the formation energies of the transition state and products, being state functions, remain unchanged. Additionally, the temperature impacts the formation energy only by a few millielectronvolts,[11,12] which is negligible compared to $\phi_{TH}$ (≈4.4 eV). Therefore, $\phi_{TH}$ can be applied universally across various material systems (with a band gap), regardless of the hydrogen environment and temperature.

In a recent publication,[13] we investigated the reduction (denitridation) of TMNs in a hydrogen radical (H*) environment and proposed that this depends upon whether H* binds to the TM or N atoms (hydrogenation). By expanding the model proposed by Van de Walle et al.[6,7] to TMN systems, we expect that the hydrogenation pathway in TMNs is determined by the work function of the host TMN system. This implies that, when the work function of a TMN system is lower than $\phi_{TH}$, H* favorably forms bonds with TM atoms. This prevents TMN from reducing, as no hydrogen binds to the N atoms. However, when the work function of TMN exceeds $\phi_{TH}$, the H−N bond is preferred, offering a pathway for reduction by the formation of volatile NH$_x$ species.

The thermodynamic feasibility of forming NH$_x$ species then governs the denitridation of TMN, which can be calculated on the basis of the change in the Gibbs free energy ($\Delta G$) for the TMN denitridation reaction (TMN + $x$H → TM + NH$_x$). Due to the formation of volatile NH$_x$ species, N vacancies form at the surface. At elevated temperatures, subsurface N atoms can then diffuse to the surface, filling the vacancies and leading to further denitridation. However, as the (electronegative) N atoms are removed from the TMN system, the work function progressively decreases, ultimately reaching $\phi_{TH}$. Consequently, the reduction reaction is self-limiting and eventually stops.

In this work, we demonstrate experimentally that the reduction of TMNs leads to a decrease in their work function. We find that the reduction reaction effectively stops when the work function drops to 4.3 ± 0.4 eV ($\phi_{TH}$), even though N atoms are still present at the surface level and a further reduction reaction is thermodynamically feasible.

We expose 5 ± 0.5 nm thin films of TiN, TaN, and NbN with a 1−2 nm surface oxynitride (TMO$_x$N$_y$) layer to H* ($10^{21\pm1}$ H* m$^{-2}$ s$^{-1}$, impinging on the sample surface[13]) at 700 °C. These experimental conditions are relevant to, e.g., fusion reactors or EUV scanners.[2,14−17] Before H* exposure, the samples are annealed at 700 °C for 2 h in a vacuum to saturate any thermally induced process. These samples are referred to as pre-exposed (pre-exp) in the text and figures. We measure changes in the chemical composition of the samples and corresponding work functions as a function of H* exposure time via angle-resolved X-ray photoelectron spectroscopy (AR-XPS) (Methodology). To minimize uncertainties associated with the quality of XPS spectral fits, we determine the atomic % of TM, N, and O atoms in the TMN samples (excluding the TaN sample, for which the TaN$_x$/TaO$_x$N$_y$ peak in the N 1s spectra is considered when calculating the N fraction) by effectively integrating the intensities of their respective XPS spectra following Shirley background subtraction. These intensities are then scaled according to their respective Scofield sensitivity factors.[18] This quantification method is consistent with the literature.[19−22]

The TiN sample predominantly undergoes surface deoxidation upon H* exposure. The work function of the pre-exposed (0 h) TiN sample is measured to be 4.9 ± 0.3 eV (Figure 1d), which is in line with the reported work function of ambient-exposed TiN (with surface TMO$_x$N$_y$).[23] As the formation of O vacancies is energetically more favorable than N vacancies on the (O-rich) TMO$_x$N$_y$ layer (Figure S1 of the Supporting Information),[13] the pre-exposed TiN sample first shows deoxidation upon 2 h of H* exposure (Figure 1a and Figure S2 of the Supporting Information). The removal of O atoms results in a lesser attenuation of photoelectrons at the surface level. Additionally, N atoms also diffuse to the surface O vacancies. The combined effect leads to an increase in the TiN fraction over the XPS probing depth, which is apparent from both the increased intensity of the TiN doublet in the Ti 2p spectra and the N/Ti ratio (panels a and c of Figure 1). Notably, the change in the full width at half maximum (FWHM) of the TiN doublet in the Ti 2p spectra is negligible, suggesting that the TiN fraction in the sample remained stable upon 2 h of H* exposure (Figure 1a). Furthermore, a slight shift in the TiN/TMO$_x$N$_y$ peak in the N 1s spectra toward a higher binding energy is due to the drop in the O fraction (Figure 1b). Because of the surface deoxidation, followed by diffusion of subsurface N atoms to the surface vacancies, the surface stoichiometry of the sample changes from approximately TiO$_{0.87}$N$_{0.85}$ to TiO$_{0.77}$N$_{0.94}$ (measured at $\theta$ = 71.75°, surface level). The removal of O atoms (more electronegative than N atoms) leads to a drop in the work function of the sample by 0.7 ± 0.2 eV (Figure 1d and Figure S8 of the Supporting Information).

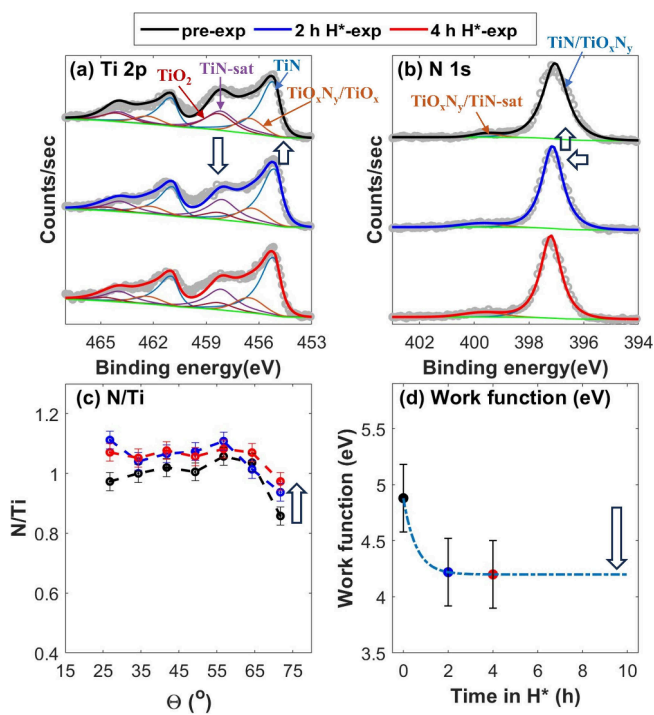

**Figure 1.** (a and b) XPS spectra of the pre-exposed (black), 2 h H*-exposed (blue), and 4 h H*-exposed (red) TiN sample, taken at a takeoff angle ($\theta$) = 34.25° along with (c) variation in the N/Ti ratio over the range of AR-XPS measurements and (d) measured work functions before and after H* exposures. Surface deoxidation during 2 h of H* exposure results in an increase in the TiN content in the XPS probing depth and a decrease in the work function to 4.2 ± 0.3 eV. No significant change in the XPS spectra and the N/Ti ratio after 2 h of H* exposure indicates that TiN is non-reducible under the performed experimental conditions. The non-reducibility of TiN is attributed to its low work function.

The work function of the sample after 2 h of H* exposure is 4.2 ± 0.3 eV (Figure 1d), which is close to the work function reported for pristine TiN.[8] Following 2 h of H* exposure, the reduction reaction effectively stops, as evident from the stabilization of the stoichiometry of the sample (Figure 1). This is likely due to unfavorable H* adsorption on surface N atoms owing to the low work function of the sample, as there are N atoms present at the surface level (Figure 1c), and further reduction of the sample is thermodynamically feasible (Figure S1 of the Supporting Information). Note that the unfavorable H* adsorption on N atoms aligns with the experimental evidence by Kura et al.[8] and simulations by Van de Walle et al.[6,7]

In contrast to the TiN sample, which predominantly shows surface deoxidation, the TaN and NbN samples do show denitridation. The work functions of the pre-exposed (0 h) TaN and NbN samples are 4.7 ± 0.3 and 5.3 ± 0.3 eV, respectively (Figures 2d and 3d). These values are in close agreement with the reported work functions of ambient-exposed TaN and NbN thin films with surface TMO$_x$N$_y$.[23] The formation of H$_2$O is energetically favorable over NH$_3$ on O-rich TaO$_x$N$_y$ and NbO$_x$N$_y$ (Figure S1 of the Supporting Information),[13] which results in the deoxidation of the surfaces. The deoxidation of the samples is evident from the decrease in the intensity of $Ta_2O_5$ and $Nb_2O_5$ doublets in the core-level TM XPS spectra and the O/TM ratio after 2 h of H* exposure (Figures 2a and 3a and Figures S3 and S4 of the Supporting Information). Because the work functions of pristine TaN and NbN are higher than $\phi_{TH}$ (>4.5 eV),[24] denitridation of the samples readily starts after deoxidation. This is evident from the shift in the $TaN_x$ and $NbN_x$ doublets in the core-level TM XPS spectra by ≈0.4 and ≈0.3 eV lower binding energies, respectively, along with an increase in the FWHM of the doublets (Figures 2a and 3a). Consistently, the N/TM ratios also decreased (Figures 2c and 3c). Furthermore, owing to the deoxidation of the surface, $TaN/TaO_xN_y$ and $NbN/NbO_xN_y$ peaks in the N 1s spectra are shifted by 0.4 eV higher binding energy (Figures 2b and 3b). Due to the reduction, the surface stoichiometry of the samples shifts from approximately $TaO_{0.87}N_{0.60}$ and $NbO_{0.47}N_{0.75}$ to $TaO_{0.42}N_{0.49}$ and $NbO_{0.23}N_{0.62}$ (measured at $\theta = 71.75°$). Additionally, the work functions of the TaN and NbN samples drop by $0.4 \pm 0.2$ and $0.5 \pm 0.2$ eV, respectively, upon 2 h of H* exposure (Figures 2d and 3d and Figure S8 of the Supporting Information).

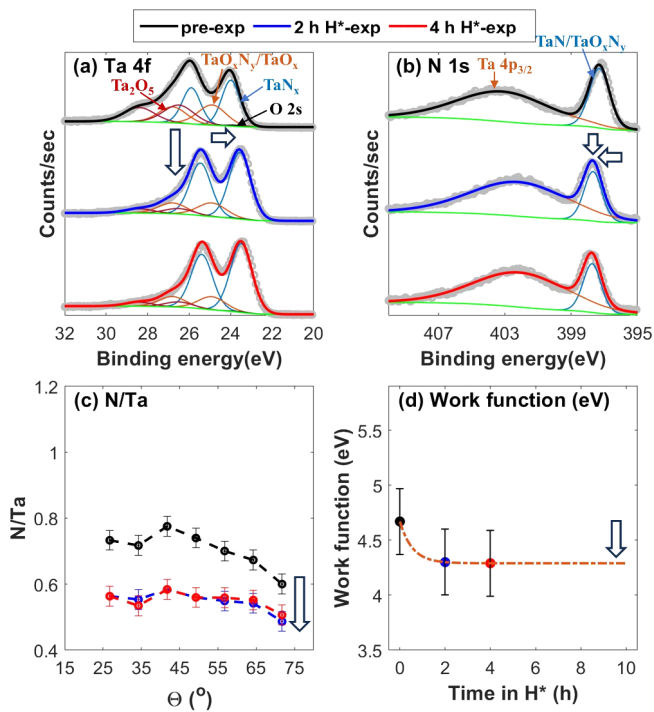

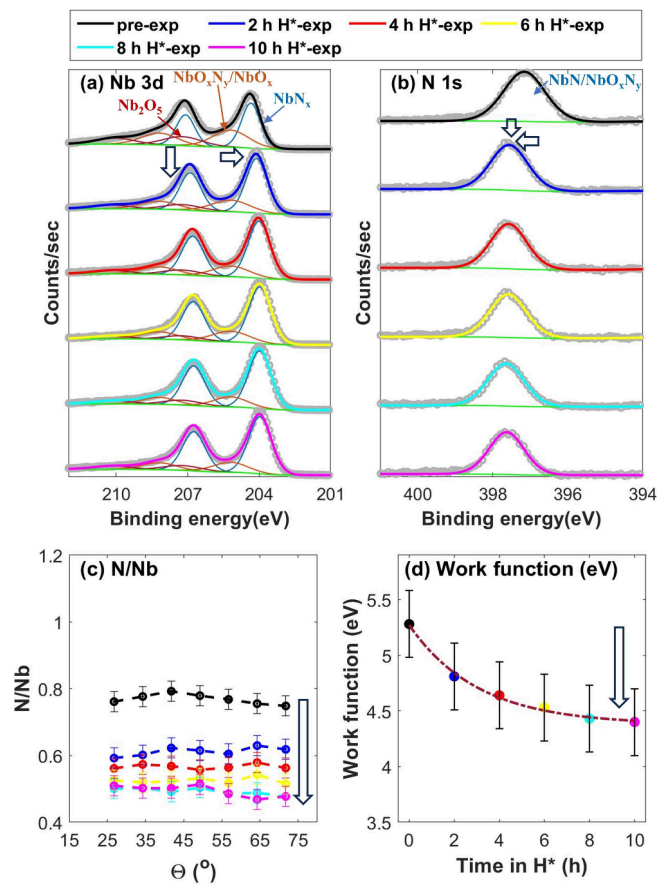

**Figure 2.** (a and b) XPS spectra of the pre-exposed (black), 2 h H*-exposed (blue), and 4 h H*-exposed (red) TaN sample, taken at $\theta = 34.25°$ along with (c) variation in the N/Ta ratio over the range of AR-XPS measurements and (d) measured work functions before and after H* exposures. Surface deoxidation along with denitridation is observed upon 2 h of H* exposure. Following 2 h of H* exposure, the work function of the sample is measured to be $4.3 \pm 0.3$ eV. Because of the low work function, the sample is not reduced any further.

The work function of the TaN sample after 2 h of H* exposure is $4.3 \pm 0.3$ eV (Figure 2d), and no further reduction of the sample is observed (Figure 2). However, the NbN sample continues denitridation over 8 h of H* exposure until its work function also reaches $4.4 \pm 0.3$ eV (Figure 3). Sufficient N atoms are present at the surfaces of the TaN and NbN samples (Figures 2c and 3c), and further formation of $NH_x$ species is thermodynamically feasible (Figure S1 of the

**Figure 3.** XPS spectra of the pre-exposed (black), 2 h H*-exposed (blue), 4 h H*-exposed (red), 6 h H*-exposed (yellow), 8 h H*-exposed (cyan), and 10 h H*-exposed (magenta) NbN sample, taken at a $\theta = 34.25°$ along with (c) variation in the N/Nb ratio over the range of AR-XPS measurements and (d) measured work functions before and after H* exposures. Reduction of the sample is observed until its work function dropped to $4.4 \pm 0.3$ eV.

Supporting Information). Therefore, the non-reducibility of the samples once their work function reaches $4.3 \pm 0.4$ eV suggests that work function directly influences their reduction in H*.

In summary, the TiN sample undergoes surface deoxidation (Figure 4a). Due to the removal of O atoms, the work function of the TiN sample drops to $4.2 \pm 0.3$ eV (Figure 4b). Notably, no further change in the chemical composition of the sample is observed afterward (Figure 4). In contrast, the TaN and NbN samples undergo denitridation, besides deoxidation (Figure 4a). However, denitridation of the TaN and NbN samples effectively stops as their work functions drop to $4.3 \pm 0.3$ and $4.4 \pm 0.3$ eV, respectively (Figure 4). The insignificant variation in measured $\phi_{TH}$ for the studied TMN samples indicates a strong correlation between the work function and reducibility of TMNs in H*. This also suggests that $\phi_{TH}$ is largely material-independent, aligning with the literature.[6,7]

In conclusion, we show that the reduction of a TMN system in H* effectively stops when its work function drops to $4.3 \pm 0.4$ eV. The value strikingly aligns with the reported $4.4 \pm 0.2$ eV work function, where $H^+$ and $H^-$ in semiconductors and insulators have equal formation energies, as proposed in refs 6 and 7. Thus, we propose that the reduction reactions of TMNs depend upon whether H* forms bonds with TM or N atoms (hydrogenation), determined by the work function of the host material. Furthermore, we propose that, when the work function drops to a threshold ($\phi_{TH}$), due to chemical alteration of the surface, H* preferably binds to the TM atoms rather than N atoms. This effectively stops the reduction reaction of the TMNs, showing a stable TMN surface stoichiometry, as no volatile species form from binding H* to the TM atoms. We propose that this model holds for a wider range of transition metal compounds (TMX, where X = O or C), using the work function as a key parameter for predicting materials' stability for hydrogen-protective coatings.

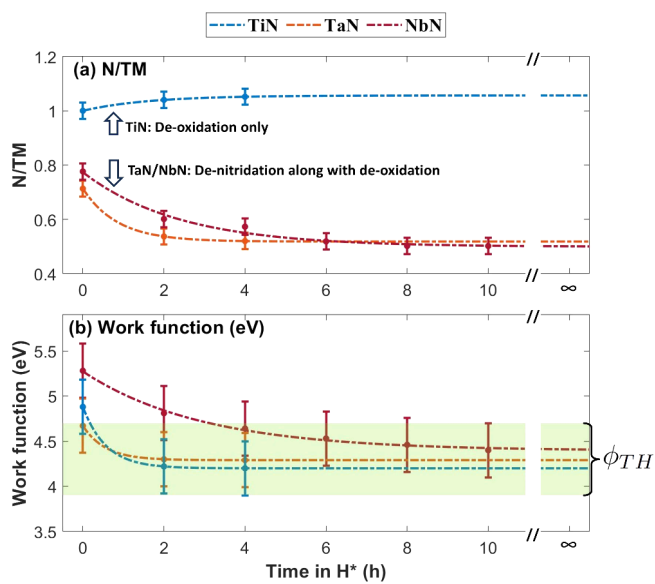

**Figure 4.** (a) N/TM ratio in the samples measured at $\theta = 34.25°$, before and after H* exposures and (b) corresponding measured work functions of the TiN (blue), TaN (orange), and NbN (maroon) samples. The N/Ti ratio increases after 2 h of H* exposure due to deoxidation of surface $TMO_xN_y$. The work function of the 2 h H*-exposed TiN sample is $4.2 \pm 0.3$ eV, and no further change in the stoichiometry of the sample is noted. In contrast, besides deoxidation, denitridation of the TaN and NbN samples occurs until their work functions drop to $4.3 \pm 0.3$ and $4.4 \pm 0.3$ eV, respectively. The fitted exponential curves show that the TMN reduction reaction effectively stops as the work function approaches $\phi_{TH}$.

**Methodology.** TiN, TaN, and NbN thin films are deposited via reactive direct current (DC) magnetron sputtering onto Si(100) substrates. The base pressure of the deposition chamber is in the low $10^{-8}$ mbar range for all of the depositions. Ar (99.999%) and $N_2$ (99.999%) with a flow rate of 15 standard cubic centimeters per minute (sccm) are used as sputtering gases. The working pressure during the depositions is $10^{-3}$ mbar. The $5 \pm 0.5$ nm TMN thin films are deposited, where the thickness is controlled by deposition time and the deposition rates were in advance calibrated via X-ray reflectivity measurements. The measurements are performed using a Malvern Panalytical Empyrean laboratory diffractometer, which uses monochromatic Cu K$\alpha_1$ radiation. The thickness of the films is chosen such that the full depth of the films is probed by AR-XPS. AR-XPS measurements are performed using a Thermo Fisher theta probe angle-resolved X-ray photoelectron spectrometer, which uses a monochromatic Al K$\alpha$ radiation source with a spot size of $400 \times 400$ μm. After depositions, the samples are transferred through air to the load lock of the exposure and XPS vacuum chambers. The exposure to ambient is minimized (<0.5 h) to limit surface oxidation (Figure 5a).

From the load lock, the samples are transported to the exposure chamber via a vacuum of low $10^{-9}$ mbar. In the exposure chamber, the samples are then first annealed at 700 °C. The base pressure of the exposure chamber is in the low $1 \times 10^{-8}$ mbar range. The samples are annealed at 700 °C for 2 h with a base pressure of low $10^{-7}$ mbar during annealing. The sample temperature is measured via a N-type thermocouple, which is clamped on the sample surface. After annealing, the samples are cooled to approximately 100 °C and then vacuum-transferred to the AR-XPS chamber (low $10^{-9}$ mbar). The corresponding AR-XPS measurements are referred to as "pre-exposed" (or pre-exp) in the text and figures (Figure 5b).

The pre-exposed samples are transferred back to the exposure chamber *in vacuo* for H* exposure. H* in the chamber is generated by thermally cracking $H_2$ with a W filament heated to ≈2000 °C. The samples are placed ≈5 cm from the cracking filament. The working pressure was set at 0.02 mbar. The samples are exposed to H* at 700 °C for 2 h. After the H* exposure, samples are cooled to about 100 °C and transferred to the AR-XPS chamber via vacuum. The corresponding AR-XPS measurements are referred to as "2 h H*-exposed" (2 h H*-exp) in the text and figures (Figure 5c).

To assess further denitridation as a function of H* exposure, the samples are repeatedly exposed to H* in a similar manner (700 °C for 2 h) until no significant change in the chemical composition of the samples is observed (panels d and e of Figure 5). AR-XPS measurements are performed on the samples after each exposure, which are labeled after the total H* exposure time.

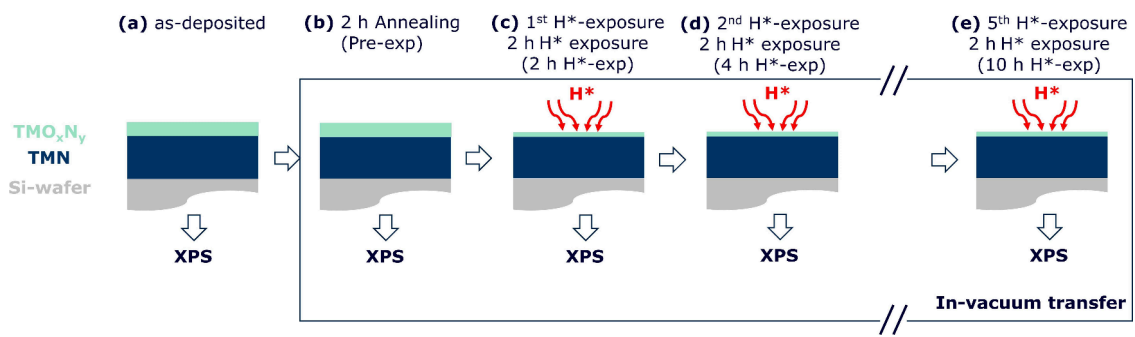

**Figure 5.** Schematic of methodology. (a) TMN samples are deposited via reactive DC magnetron sputtering. The 1−2 nm thin $TMO_xN_y$ layers form on the samples' surfaces during ambient transfer. (b) TMN samples are annealed at 700 °C. (c−e) Samples are repeatedly exposed to H* at 700 °C for 2 h until no further change in their stoichiometry is observed. AR-XPS measurements are performed on the samples after each process (annealing/H* exposure).

Changes in the stoichiometry of the samples are evaluated on the basis of the core-level TM, N 1s, and O 1s AR-XPS spectra. Core-level TM and N 1s XPS spectra taken at $\theta$ (from the surface normal) of 34.25° are discussed in detail in the text. The spectra are fitted with Voigt profile doublets/peaks, following Shirley background correction (the fitting method is described in the Supporting Information). A comparison between core-level TM, N 1s, O 1s, and Si 2p XPS spectra taken over the range of AR-XPS measurements ($\theta$ ranging from 26.75° to 71.75°, providing probing depth from ≈1.5 to ≈5 nm) before and after H* exposures is also provided in Figures S5−S7 of the Supporting Information. To quantify N and O atomic losses, the ratios between atomic % of N and TM (N/TM) and atomic % of O and TM (O/TM) are calculated over the range of AR-XPS measurements. Changes in the N/TM ratios are discussed in the text, while O/TM ratios are provided in Figures S2−S4 of the Supporting Information.

The work function of the samples is measured via XPS in normal lens mode.[25] A negative bias of 16.4 V is applied with the valence band (VB) and low kinetic energy (LKE) spectra. To account for this bias, we also collected VB spectra without any applied bias. Because TMN samples are sufficiently conductive, the Fermi level of the unbiased samples is set as the reference point at zero binding energy. To ensure consistency, the binding energy of the biased sample is adjusted to align with the Fermi level of the unbiased sample (LKE and VB spectra of the samples are provided in Figure S8 of the Supporting Information). In our instrument, the sample plane is not parallel to the entrance of the electron analyzer, which leads to a systematic deviation between the measured and actual work function according to refs 25 and 26. By comparing the measured and reported work functions of sputter-cleaned polycrystalline foils of Au, Cu, and Ag (Table S5 and Figures S9 and S10 of the Supporting Information), we determined this systematic offset in the measured work function to be $-1.0 \pm 0.2$ eV. The work function values presented in the text and figures have already been adjusted to account for this offset. Note that the uncertainty associated with the measured secondary electron cutoff is $\pm 0.1$ eV. Therefore, the certainty in the difference between the measured work functions is only $\pm 0.2$ eV.

## ■ ASSOCIATED CONTENT

### ⓈⓘSupporting Information

The Supporting Information is available free of charge at https://pubs.acs.org/doi/10.1021/acs.jpclett.4c02259.

$\Delta G$ for the reduction reactions of TMN$_x$ and TMO$_y$, O/TM ratios, comparison between the core-level TM, N 1s, O 1s, and Si 2p XPS spectra as a function of $\theta$, methodology used for fitting the XPS spectra, LKE and VB spectra, XPS survey spectra of Au, Cu, and Ag foils, and calculated offset in the measured work function (PDF)

## ■ AUTHOR INFORMATION

### Corresponding Author

Abdul Rehman − Industrial Focus Group XUV Optics, MESA+ Institute for Nanotechnology, University of Twente, 7522NB Enschede, Netherlands; orcid.org/0000-0002-8846-9975; Email: a.rehman@utwente.nl

### Authors

Robbert W. E. van de Kruijs − Industrial Focus Group XUV Optics, MESA+ Institute for Nanotechnology, University of Twente, 7522NB Enschede, Netherlands; orcid.org/0000-0002-4738-0819

Wesley T. E. van den Beld − Industrial Focus Group XUV Optics, MESA+ Institute for Nanotechnology, University of Twente, 7522NB Enschede, Netherlands; orcid.org/0000-0002-5449-3838

Jacobus M. Sturm − Industrial Focus Group XUV Optics, MESA+ Institute for Nanotechnology, University of Twente, 7522NB Enschede, Netherlands; orcid.org/0000-0002-0731-6329

Marcelo Ackermann − Industrial Focus Group XUV Optics, MESA+ Institute for Nanotechnology, University of Twente, 7522NB Enschede, Netherlands

Complete contact information is available at:
https://pubs.acs.org/10.1021/acs.jpclett.4c02259

### Notes

The authors declare no competing financial interest.

## ■ ACKNOWLEDGMENTS

This work has been carried out in the frame of the Industrial Partnership Program "X-tools", Project 741.018.301, funded by the Netherlands Organization for Scientific Research, ASML, Carl Zeiss SMT, and Malvern Panalytical. The authors acknowledge the support of the Industrial Focus Group XUV Optics at the MESA+ Institute for Nanotechnology at the University of Twente. The authors extend their gratitude to Stefan van Vliet from the University of Twente for his invaluable insights and discussions on XPS.

## ■ REFERENCES

(1) Hino, T.; Akiba, M. Japanese developments of fusion reactor plasma facing components. *Fusion Eng. Des.* 2000, 49, 97−105.

(2) Nemanič, V. Hydrogen permeation barriers: Basic requirements, materials selection, deposition methods, and quality evaluation. *Nucl. Mater. Energy* 2019, 19, 451−457.

(3) Boudghene Stambouli, A; Traversa, E Fuel cells, an alternative to standard sources of energy. *Renewable Sustainable Energy Rev.* 2002, 6, 295−304.

(4) Zhang, F.; Zhao, P.; Niu, M.; Maddy, J. The survey of key technologies in hydrogen energy storage. *Int. J. Hydrogen Energy* 2016, 41, 14535−14552.

(5) Matějíček, J.; Veverka, J.; Nemanič, V.; Cvrček, L.; Lukáč, F.; Havránek, V.; Illková, K. Characterization of less common nitrides as potential permeation barriers. *Fusion Eng. Des.* 2019, 139, 74−80.

(6) Van de Walle, C. G.; Neugebauer, J. Universal alignment of hydrogen levels in semiconductors, insulators and solutions. *Nature* 2003, 423, 626−628.

(7) Van de Walle, C. G. Universal alignment of hydrogen levels in semiconductors and insulators. *Phys. B* 2006, 376, 1−6.

(8) Kura, C.; Kunisada, Y.; Tsuji, E.; Zhu, C.; Habazaki, H.; Nagata, S.; Müller, M. P.; De Souza, R. A.; Aoki, Y. Hydrogen separation by nanocrystalline titanium nitride membranes with high hydride ion conductivity. *Nat. Energy* 2017, 2, 786−794.

(9) Kura, C.; Fujimoto, S.; Kunisada, Y.; Kowalski, D.; Tsuji, E.; Zhu, C.; Habazaki, H.; Aoki, Y. Enhanced hydrogen permeability of hafnium nitride nanocrystalline membranes by interfacial hydride conduction. *J. Mater. Chem. A* 2018, 6, 2730−2741.

(10) Saito, M.; Kura, C.; Toriumi, H.; Hinokuma, S.; Ina, T.; Habazaki, H.; Aoki, Y. Formation of mobile hydridic defects in zirconium nitride films with n-type semiconductor properties. *ACS Appl. Electron. Mater.* **2021**, *3*, 3980−3989.

(11) Chase, M. W. *NIST-JANAF Thermochemical Tables*, 4th ed.; American Institute of Physics (AIP): College Park, MD, 1998; pp 1529−1564.

(12) Barin, I.; Platzki, G. *Thermochemical Data of Pure Substances*; Wiley Online Library: Hoboken, NJ, 1989; Vol. *304*.

(13) Rehman, A.; van de Kruijs, R. W.; van den Beld, W. T.; Sturm, J. M.; Ackermann, M. Chemical interaction of hydrogen radicals (H*) with transition metal nitrides. *J. Phys. Chem. C* **2023**, *127*, 17770−17780.

(14) Brouns, D.; Bendiksen, A.; Broman, P.; Casimiri, E.; Colsters, P.; Delmastro, P.; de Graaf, D.; Janssen, P.; van de Kerkhof, M.; Kramer, R.; et al. NXE pellicle: Offering a EUV pellicle solution to the industry. *Proceedings of the SPIE Advanced Lithography*; San Jose, CA, Feb 21−25, 2016; Extreme Ultraviolet (EUV) Lithography VII, Vol. *9776*, p 97761Y, DOI: 10.1117/12.2221909.

(15) Luo, L.-M.; Liu, Y.-L.; Liu, D.-G.; Zheng, L.; Wu, Y.-C. Preparation technologies and performance studies of tritium permeation barriers for future nuclear fusion reactors. *Surf. Coat. Technol.* **2020**, *403*, 126301.

(16) van de Kerkhof, M.; Yakunin, A. M.; Kvon, V.; Nikipelov, A.; Astakhov, D.; Krainov, P.; Banine, V. EUV-induced hydrogen plasma and particle release. *Radiat. Eff. Defects Solids* **2022**, *177*, 486−512.

(17) Zoldesi, C.; Bal, K.; Blum, B.; Bock, G.; Brouns, D.; Dhalluin, F.; Dziomkina, N.; Espinoza, J. D. A.; de Hoogh, J.; Houweling, S.; et al. Progress on EUV pellicle development. *Proceedings of the SPIE Advanced Lithography*; San Jose, CA, Feb 23−27, 2014; Extreme Ultraviolet (EUV) Lithography V, Vol. *9048*, p 90481N, DOI: 10.1117/12.2049276.

(18) Scofield, J. H. Hartree-Slater subshell photoionization cross-sections at 1254 and 1487 eV. *J. Electron Spectrosc. Relat. Phenom.* **1976**, *8*, 129−137.

(19) Greczynski, G.; Hultman, L. X-ray photoelectron spectroscopy: Towards reliable binding energy referencing. *Prog. Mater. Sci.* **2020**, *107*, 100591.

(20) Lamour, P.; Fioux, P.; Ponche, A.; Nardin, M.; Vallat, M.-F.; Dugay, P.; Brun, J.-P.; Moreaud, N.; Pinvidic, J.-M. Direct measurement of the nitrogen content by XPS in self-passivated TaN$_x$ thin films. *Surf. Interface Anal.* **2008**, *40*, 1430−1437.

(21) Zier, M.; Oswald, S.; Reiche, R.; Wetzig, K. XPS and ARXPS investigations of ultra thin TaN films deposited on SiO$_2$ and Si. *Appl. Surf. Sci.* **2005**, *252*, 234−239.

(22) Havey, K.; Zabinski, J.; Walck, S. The chemistry, structure, and resulting wear properties of magnetron-sputtered NbN thin films. *Thin Solid Films* **1997**, *303*, 238−245.

(23) Fujii, R.; Gotoh, Y.; Liao, M.; Tsuji, H.; Ishikawa, J. Work function measurement of transition metal nitride and carbide thin films. *Vacuum* **2006**, *80*, 832−835.

(24) Patsalas, P.; Kalfagiannis, N.; Kassavetis, S.; Abadias, G.; Bellas, D.; Lekka, C.; Lidorikis, E. Conductive nitrides: Growth principles, optical and electronic properties, and their perspectives in photonics and plasmonics. *Mater. Sci. Eng., R* **2018**, *123*, 1−55.

(25) Kim, J. W.; Kim, A. Absolute work function measurement by using photoelectron spectroscopy. *Curr. Appl. Phys.* **2021**, *31*, 52−59.

(26) Helander, M.; Greiner, M.; Wang, Z.; Lu, Z. Pitfalls in measuring work function using photoelectron spectroscopy. *Appl. Surf. Sci.* **2010**, *256*, 2602−2605.